%% file: main.tex
\documentclass[11pt]{article}
\usepackage[dvips]{graphicx}
\usepackage{epsfig}
\usepackage[left]{lineno}
\usepackage{amsmath}
\usepackage{amssymb}
\usepackage{color}
\usepackage{hyperref}
\usepackage{adjustbox}
\usepackage{xspace}
\usepackage{multirow}
\usepackage{easyReview}
\usepackage{booktabs}
\usepackage{xltabular}
\usepackage{relsize}
\usepackage{enumitem}
\usepackage{alphalph}
\usepackage{calc}
\usepackage{float}
\usepackage{cite}
\usepackage{orcidlink}
\usepackage{geometry}
\usepackage[separate-uncertainty,bracket-numbers = false,
table-align-uncertainty = true, table-align-exponent=false, 
exponent-product=\cdot,parse-numbers = false]{siunitx}
\DeclareSIUnit{\GeVc}{GeV/\textit{c}}
\DeclareSIUnit{\mmGeV}{GeV^2/\textit{c}^4}
\DeclareSIUnit{\MeVc}{MeV/\textit{c}}

\definecolor{darkgreen}{rgb}{0.23, 0.5, 0.23}
\definecolor{amber}{rgb}{0.83, 0.67, 0}

\catcode`\@=11 
\def\parenbar{\mathpalette\p@renb@r}
\def\p@renb@r#1#2{\vbox{%
		\ifx#1\scriptscriptstyle \dimen@.7em\dimen@ii.2em\else
		\ifx#1\scriptstyle \dimen@.8em\dimen@ii.25em\else
		\dimen@1em\dimen@ii.4em\fi\fi \offinterlineskip
		\ialign{\hfill##\hfill\cr
			\vbox{\hrule width\dimen@ii}\cr
			\noalign{\vskip-.3ex}%
			\hbox to\dimen@{$\mathchar300\hfil\mathchar301$}\cr
			\noalign{\vskip-.3ex}%
			$#1#2$\cr}}}
\catcode`\@=12 
\def\nuan{\ensuremath{\parenbar{\nu}\kern-0.4ex}}

\def\PiZ{\ensuremath{\pi^{0}}}
\def\Pi{\ensuremath{\pi^{\pm}}}
\def\PiP{\ensuremath{\pi^{+}}}
\def\PiM{\ensuremath{\pi^{-}}}
\def\Mu{\ensuremath{\mu^{\pm}}}

\def\MuP{\ensuremath{\mu^{+}}}
\def\MuM{\ensuremath{\mu^{-}}}
\def\K{\ensuremath{K^{\pm}}}
\def\KP{\ensuremath{K^{+}}}
\def\NuANuMu{\ensuremath{\nuan_\mu}}
\def\NuANuE{\ensuremath{\nuan_e}}

\def\NuMu{\ensuremath{\nu_\mu}}

\def\PiMuNu{\ensuremath{\Pi \to \Mu \NuANuMu}}

\def\KMuNu{\ensuremath{\K \to \Mu \NuANuMu}}
\def\KMuNuP{\ensuremath{\KP \to \MuP \NuMu}}

\def\KPiENu{\ensuremath{\K \to \PiZ e^\pm \NuANuE}}

\def\KPiPi{\ensuremath{\KP \to \PiP \PiZ}}

\def\KPiNuNuP{\ensuremath{\KP \to \PiP \nu \bar\nu}}
\def\SqMmiss{\ensuremath{m^2_{\rm{miss}}}}

\def\sig{\ensuremath{{\rm{sig}}}}
\def\norm{\ensuremath{{\rm{norm}}}}
\def\int{\ensuremath{{\rm{int}}}}
\def\stat{\ensuremath{{\rm{stat}}}}
\def\syst{\ensuremath{{\rm{syst}}}}
\def\LKr{\ensuremath{{\rm{LKr}}}}
	
\def\geant {\mbox{\textsc{Geant4}}\xspace}
\def\genie {\mbox{\textsc{Genie}}\xspace}

\newgeometry{left=1.1in,right=1.1in,top=0.6in,bottom=0.7in}

\let\oldequation\equation
\let\oldendequation\endequation
\renewenvironment{equation}
    {\linenomathNonumbers\oldequation}
    {\oldendequation\endlinenomath}

\newcommand{\appropto}{\mathrel{\vcenter{
  \offinterlineskip\halign{\hfil$##$\cr
    \propto\cr\noalign{\kern2pt}\sim\cr\noalign{\kern-2pt}}}}}

\begin{document}
\centerline{\LARGE EUROPEAN ORGANIZATION FOR NUCLEAR RESEARCH}

\vspace{10mm}
{\flushright{
 CERN-EP-2024-324\\
  December 4, 2024\\
  revised January 20, 2025\\
}}
\vspace{20mm}

\begin{center}
\boldmath
{\bf \Large First detection of a tagged neutrino in the NA62 experiment}

\unboldmath
\end{center}

\vspace{5mm}
\begin{center}
{\Large The NA62 Collaboration}
\footnote{Corresponding authors: Bianca De Martino, Mathieu Perrin-Terrin \\
email : \url{bianca.de.martino@cern.ch}, 
\url{mathieu.perrin-terrin@cern.ch} }\\
\end{center}

\begin{abstract}
The NA62 experiment at the CERN SPS reports the first detection of a tagged neutrino candidate based on the data collected in 2022. The candidate consists of a $K^+ \rightarrow \mu^+ \nu_\mu$ decay where the charged particles are reconstructed
and the neutrino is detected through a charged-current interaction in a liquid krypton calorimeter.
\end{abstract}

\vspace{20mm}

\begin{center}
    \textit{Accepted for publication in Physics Letters B}
\end{center}

\newpage
\input{authors_nutagv5}

\clearpage
\input{body}
\section*{Acknowledgments}
\input{acknowrun2022_optC_nutag-v3}

\bibliographystyle{elsarticle-num}
\bibliography{ref.bib}

\end{document}

%% file: authors_nutagv5.tex
\newcommand{\orcimg}{\raisebox{-0.3\height}{\includegraphics[height=\fontcharht\font`A]{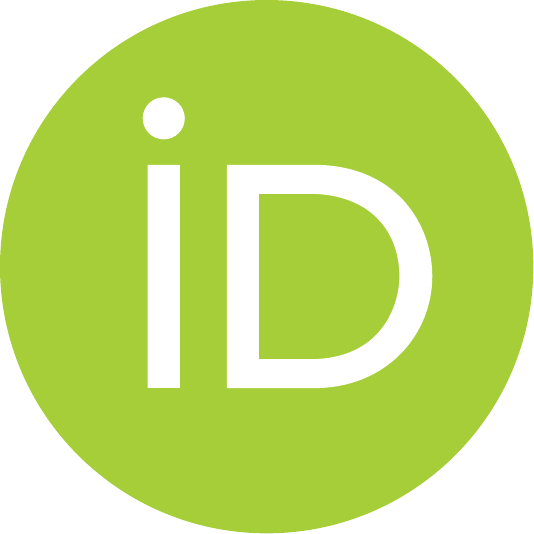}}}
\newcommand{\orcid}[1]{\href{https://orcid.org/#1}{\orcimg}}

\centerline{\bf The NA62 Collaboration}
\vspace{2.0cm}
%
%

\begin{raggedright}
\noindent
{\bf Universit\'e Catholique de Louvain, Louvain-La-Neuve, Belgium}\\
 E.~Cortina Gil\orcid{0000-0001-9627-699X},
 J.~Jerhot$\,${\footnotemark[1]}\orcid{0000-0002-3236-1471},
 N.~Lurkin\orcid{0000-0002-9440-5927}
\vspace{0.5cm}

{\bf TRIUMF, Vancouver, British Columbia, Canada}\\
 T.~Numao\orcid{0000-0001-5232-6190},
 B.~Velghe\orcid{0000-0002-0797-8381},
 V. W. S.~Wong\orcid{0000-0001-5975-8164}
\vspace{0.5cm}

{\bf University of British Columbia, Vancouver, British Columbia, Canada}\\
 D.~Bryman$\,${\footnotemark[2]}\orcid{0000-0002-9691-0775}
\vspace{0.5cm}

{\bf Charles University, Prague, Czech Republic}\\
 Z.~Hives\orcid{0000-0002-5025-993X},
 T.~Husek$\,${\footnotemark[3]}\orcid{0000-0002-7208-9150},
 K.~Kampf\orcid{0000-0003-1096-667X},
 M.~Kolesar\orcid{0000-0002-9085-2252},
 M.~Koval\orcid{0000-0002-6027-317X}
\vspace{0.5cm}

{\bf Aix Marseille University, CNRS/IN2P3, CPPM, Marseille, France}\\
 B.~De Martino$\,$\renewcommand{\thefootnote}{\fnsymbol{footnote}}\footnotemark[1]\renewcommand{\thefootnote}{\arabic{footnote}}\orcid{0000-0003-2028-9326},
 M.~Perrin-Terrin$\,$\renewcommand{\thefootnote}{\fnsymbol{footnote}}\footnotemark[1]\renewcommand{\thefootnote}{\arabic{footnote}}\orcid{0000-0002-3568-1956}
\vspace{0.5cm}

{\bf Max-Planck-Institut f\"ur Physik (Werner-Heisenberg-Institut), Garching, Germany}\\
 B.~D\"obrich\orcid{0000-0002-6008-8601},
 S.~Lezki\orcid{0000-0002-6909-774X},
 J.~Schubert$\,${\footnotemark[4]}\orcid{0000-0002-5782-8816}
\vspace{0.5cm}

{\bf Institut f\"ur Physik and PRISMA Cluster of Excellence, Universit\"at Mainz, Mainz, Germany}\\
 A. T.~Akmete\orcid{0000-0002-5580-5477},
 R.~Aliberti$\,${\footnotemark[5]}\orcid{0000-0003-3500-4012},
 M.~Ceoletta$\,${\footnotemark[6]}\orcid{0000-0002-2532-0217},
 L.~Di Lella\orcid{0000-0003-3697-1098},
 N.~Doble\orcid{0000-0002-0174-5608}, 
 L.~Peruzzo\orcid{0000-0002-4752-6160},
 S.~Schuchmann\orcid{0000-0002-8088-4226},
 H.~Wahl\orcid{0000-0003-0354-2465},
 R.~Wanke\orcid{0000-0002-3636-360X}
\vspace{0.5cm}

{\bf Dipartimento di Fisica e Scienze della Terra dell'Universit\`a e INFN, Sezione di Ferrara, Ferrara, Italy}\\
 P.~Dalpiaz,
 R.~Negrello\orcid{0009-0008-3396-5550},
 I.~Neri\orcid{0000-0002-9669-1058},
 F.~Petrucci\orcid{0000-0002-7220-6919},
 M.~Soldani\orcid{0000-0003-4902-943X}
% ...
\vspace{0.5cm}

{\bf School of Physics, University of Bristol, Bristol, United Kingdom}\\
 H.~Heath\orcid{0000-0001-6576-9740}
\vspace{0.5cm}

{\bf School of Physics and Astronomy, University of Glasgow, Glasgow, United Kingdom}\\
 D.~Britton\orcid{0000-0001-9998-4342},
 A.~Norton\orcid{0000-0001-5959-5879},
 D.~Protopopescu\orcid{0000-0002-8047-6513}
\vspace{0.5cm}

{\bf Physics Department, University of Lancaster, Lancaster, United Kingdom}\\
 J. B.~Dainton,
 L.~Gatignon\orcid{0000-0001-6439-2945},
 R. W. L.~Jones\orcid{0000-0002-6427-3513}
\vspace{0.5cm}

{\bf Department of Physics and Astronomy, George Mason University, Fairfax, Virginia, USA}\\
 P.~Cooper,
 D.~Coward$\,${\footnotemark[24]}\orcid{0000-0001-7588-1779},
 P.~Rubin\orcid{0000-0001-6678-4985}
\vspace{0.5cm}

{\bf Authors affiliated with an international laboratory covered by a cooperation agreement with CERN}\\
 A.~Baeva,
 D.~Baigarashev$\,${\footnotemark[25]}\orcid{0000-0001-6101-317X},
 V.~Bautin\orcid{0000-0002-5283-6059},
 D.~Emelyanov,
 T.~Enik\orcid{0000-0002-2761-9730}, 
 V.~Falaleev$\,${\footnotemark[17]}\orcid{0000-0003-3150-2196},
 V.~Kekelidze\orcid{0000-0001-8122-5065},
 D.~Kereibay,
 A.~Korotkova,
 L.~Litov$\,${\footnotemark[8]}\orcid{0000-0002-8511-6883}, 
 D.~Madigozhin\orcid{0000-0001-8524-3455},
 M.~Misheva$\,${\footnotemark[26]},
 N.~Molokanova,
 I.~Polenkevich,
 Yu.~Potrebenikov\orcid{0000-0003-1437-4129}, 
 K.~Salamatin\orcid{0000-0001-6287-8685},
 S.~Shkarovskiy
\vspace{0.5cm}

{\bf Authors affiliated with an Institute formerly covered by a cooperation agreement with CERN}\\
 S.~Fedotov,
 K.~Gorshanov\orcid{0000-0001-7912-5962},
 E.~Gushchin\orcid{0000-0001-8857-1665},
 S.~Kholodenko$\,${\footnotemark[27]}\orcid{0000-0002-0260-6570},
 A.~Khotyantsev,
 Y.~Kudenko\orcid{0000-0003-3204-9426}, 
 V.~Kurochka,
 V.~Kurshetsov\orcid{0000-0003-0174-7336},
 A.~Mefodev,
 V.~Obraztsov\orcid{0000-0002-0994-3641},
 A.~Okhotnikov\orcid{0000-0003-1404-3522}, 
 A.~Sadovskiy\orcid{0000-0002-4448-6845},
 V.~Sugonyaev\orcid{0000-0003-4449-9993},
 O.~Yushchenko\orcid{0000-0003-4236-5115}
\vspace{0.5cm}

\end{raggedright}

%
%
\vspace{5mm}
\setcounter{footnote}{0}
\newlength{\basefootnotesep}
\setlength{\basefootnotesep}{\footnotesep}

\renewcommand{\thefootnote}{\fnsymbol{footnote}}
\noindent
$^{\footnotemark[1]}${Corresponding authors: B.~De~Martino, M.~Perrin-Terrin, \\email: bianca.de.martino@cern.ch, mathieu.perrin-terrin@cern.ch}\\
\renewcommand{\thefootnote}{\arabic{footnote}}
$^{1}${Present address: Max-Planck-Institut f\"ur Physik (Werner-Heisenberg-Institut), D-85748 Garching, Germany} \\
$^{2}${Also at TRIUMF, Vancouver, British Columbia, V6T 2A3, Canada} \\
$^{3}${Also at School of Physics and Astronomy, University of Birmingham, Birmingham, B15 2TT, UK} \\
$^{4}${Also at Department of Physics, Technical University of Munich, M\"unchen, D-80333, Germany} \\
$^{5}${Present address: Institut f\"ur Kernphysik and Helmholtz Institute Mainz, Universit\"at Mainz, Mainz, D-55099, Germany} \\
$^{6}${Also at CERN, European Organization for Nuclear Research, CH-1211 Geneva 23, Switzerland} \\
$^{7}${Also at Dipartimento di Scienze Fisiche, Informatiche e Matematiche, Universit\`a di Modena e Reggio Emilia, I-41125 Modena, Italy} \\
$^{8}${Also at Faculty of Physics, University of Sofia, BG-1164 Sofia, Bulgaria} \\
$^{9}${Present address: Scuola Superiore Meridionale e INFN, Sezione di Napoli, I-80138 Napoli, Italy} \\
$^{10}${Also at Department of Industrial Engineering, University of Roma Tor Vergata, I-00173 Roma, Italy} \\
$^{11}${Also at Department of Electronic Engineering, University of Roma Tor Vergata, I-00173 Roma, Italy} \\
$^{12}${Also at Universit\`a degli Studi del Piemonte Orientale, I-13100 Vercelli, Italy} \\
$^{13}${Present address: Universit\'e Catholique de Louvain, B-1348 Louvain-La-Neuve, Belgium} \\
$^{14}${Also at Gran Sasso Science Institute, I-67100 L'Aquila,  Italy} \\
$^{15}${Also at al-Farabi Kazakh National University, 050040 Almaty, Kazakhstan} \\
$^{16}${Also at Universidad de Guanajuato, 36000 Guanajuato, Mexico} \\
$^{17}${Present address: INFN, Sezione di Perugia, I-06100 Perugia, Italy} \\
$^{18}${Present address: Charles University, 116 36 Prague 1, Czech Republic} \\
$^{19}${Also at INFN, Sezione di Perugia, I-06100 Perugia, Italy} \\
$^{20}${Present address: Physics Department, University of Lancaster, Lancaster, LA1 4YB, UK} \\
$^{21}${Present address: Syracuse University, Syracuse, NY 13244, USA} \\
$^{22}${Present address: Laboratori Nazionali di Frascati, I-00044 Frascati, Italy} \\
$^{23}${Present address: Dipartimento di Fisica dell'Universit\`a e INFN, Sezione di Genova, I-16146 Genova, Italy} \\
$^{24}${Also at SLAC National Accelerator Laboratory, Stanford University, Menlo Park, CA 94025, USA} \\
$^{25}${Also at L. N. Gumilyov Eurasian National University, 010000 Nur-Sultan, Kazakhstan} \\
$^{26}${Present address: Institute of Nuclear Research and Nuclear Energy of Bulgarian Academy of Science (INRNE-BAS), BG-1784 Sofia, Bulgaria} \\
$^{27}${Present address: INFN, Sezione di Pisa, I-56100 Pisa, Italy} \\

%% file: body.tex
\section*{Introduction}
\label{sec:intro}
The next generation of long-baseline neutrino experiments~\cite{hep-ph_DUNE_2020, hep-ph_HyperKamiokande_2018} will collect datasets more than ten times larger than
their predecessors. Controlling systematic uncertainties in neutrino oscillation studies in these experiments becomes crucial~\cite{hep-ph_ESPP_2020}, and will be vital for the precision measurements required to probe the three-neutrino paradigm~\cite{hep-ph_HuberEtAl_2022}.
These experiments require precise knowledge of neutrino fluxes, cross-sections, interaction models, and energy 
estimators with well-controlled biases~\cite{hep-ph_HuberEtAl_2008,hep-ph_BrancaEtAl_2021,hep-ph_HyperKamiokande_2018,hep-ph_DUNE_2021,hep-ph_HuberEtAl_2022}.

A tagged-neutrino experiment~\cite{hep-ph_Pontecorvo_1979,hep-ph_Nedyalkov_1984,hep-ph_Bohm_1987,hep-ph_Bernstein_1993,hep-ph_PERRIN-TERRIN_2022} would provide improved systematic uncertainties and neutrino energy estimates. 
The setup proposed recently~\cite{hep-ph_PERRIN-TERRIN_2022} features an instrumented beamline to track charged particles. Neutrinos from \PiMuNu\ and \KMuNu\ decays are reconstructed based on energy and momentum conservation at the decay vertex.
Hence, the neutrino fluxes from \PiMuNu\ and \KMuNu\ decays, and from all other \K\ decays, particularly \KPiENu, are precisely determined. Moreover, the detected neutrinos can be individually associated with their decay parents, thus 
determining their flavour and chirality and
obtaining a precise estimate of their energy from the decay kinematics. Compared to energy reconstruction algorithms based on the measured neutrino interaction final state, which typically yield 10--30\% precision with significant biases~\cite{hep-ph_HyperKamiokande_2018,hep-ph_FriedlandEtAl_2019}, the kinematic reconstruction provides sub-percent precision and negligible bias~\cite{hep-ph_PERRIN-TERRIN_2022}.
In short-baseline experiments, neutrino tagging would allow precise measurements of the total and differential interaction cross-sections. In long-baseline experiments, tagging the neutrino interactions in the far detector
would improve the study of neutrino oscillations due to the excellent energy resolution, reduced backgrounds from non-oscillated neutrinos, and improved flavour and interaction identification.

The development of the neutrino tagging technique has been hindered so far by the prohibitive particle rates in a neutrino beamline. These obstacles are being overcome by the development of neutrino beamlines operating with slow extraction~\cite{hep-ph_ENUBET_2023, hep-ph_Baratto-RoldanEtAl_2024} and beam spectrometers capable of operating in a high rate environment~\cite{hep-ph_BorgatoEtAl_2023}.

The NA62 experiment at the CERN SPS allows the possibility of a feasibility study of the neutrino tagging technique. To measure the properties of the very rare \KPiNuNuP\ decay~\cite{CortinaGil2021}, NA62 uses a high-intensity \KP\ beam which copiously produces neutrinos, most importantly via \KMuNuP\ decays. The experiment employs a state-of-the-art beam spectrometer~\cite{hep-ph_AglieriRinellaEtAl_2019} and a liquid krypton calorimeter which can be used as an instrumented target to detect neutrino interactions. Associating neutrino interactions with the \KP\ and the \MuP\ from their parent decays would compellingly demonstrate the feasibility of the neutrino tagging technique.  

In the following, we report the first tagged neutrino detection based on data collected by NA62 in 2022. Details about the detector are given in Section~\ref{sec:beam_det} and event selection is described in Section~\ref{sec:evtsel}. Section~\ref{sec:bkg} reports the background assessments and Section~\ref{sec:signal} the signal expectation.  The properties of the tagged neutrino candidate are given in Section~\ref{sec:results}.

\section{Beam, detector and data sample}
\label{sec:beam_det}

\begin{figure}[t]
\begin{center}
\resizebox{\textwidth}{!}{\includegraphics{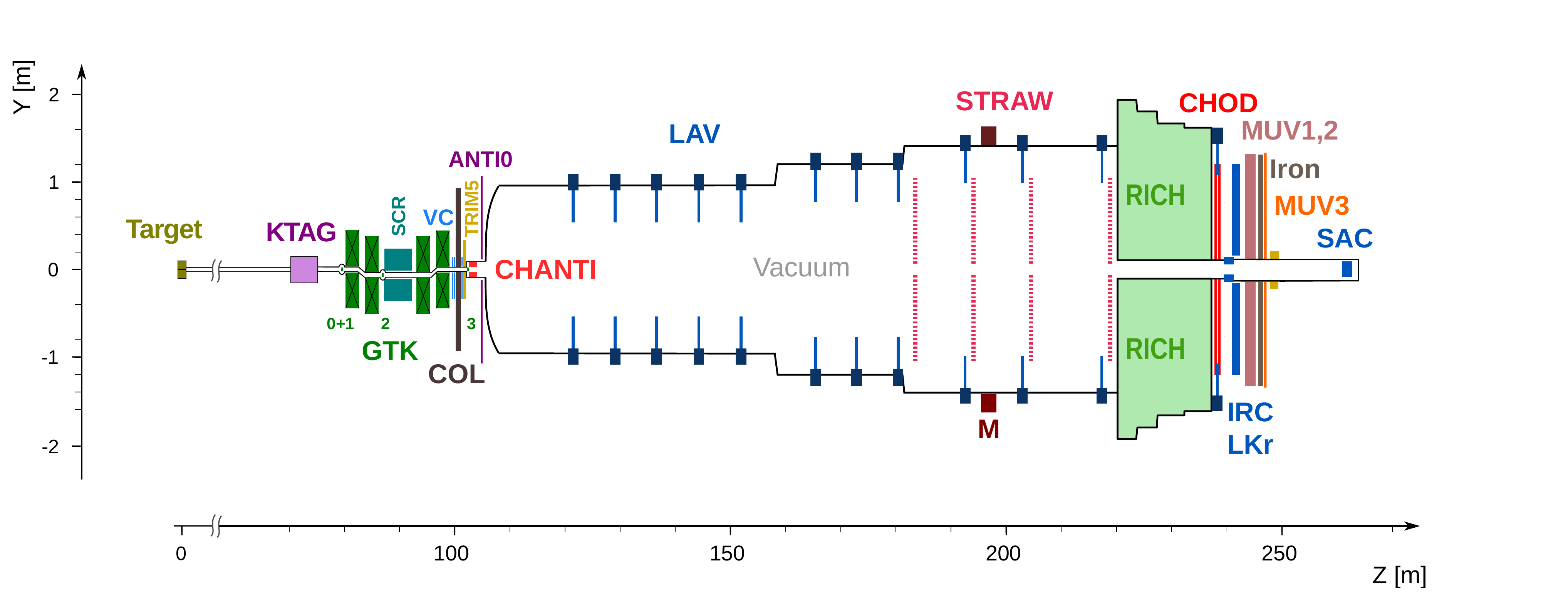}}
\end{center}
\vspace{-10mm}
\caption{Schematic side view of the NA62 beamline and detector used in 2022. Information from VC and ANTI0 counters is not used in this analysis.}
\label{fig:detector}
\end{figure}

The layout of the NA62 beamline and detector~\cite{hep-ph_NA62_2017} is shown schematically in \autoref{fig:detector}. An unseparated secondary beam of \PiP\ (70\%), protons (23\%) and \KP\ (6\%) is created by directing \SI{400}{\GeV} protons extracted from the CERN SPS onto a beryllium target in spills of \SI{4.8}{\s} duration. The target position defines the origin of the NA62 reference system: the beam travels along the $Z$ axis in the positive direction (downstream), the $Y$ axis points vertically up, and the $X$ axis is directed to form a right-handed coordinate system. The central beam momentum is \SI{75}{\GeVc}, with a momentum spread of 1\% (rms).

Beam kaons are tagged with a time resolution of \SI{70}{\ps} by a differential Cherenkov counter (KTAG), which uses nitrogen gas at \SI{1.75}{\bar} pressure contained in a \SI{5}{m} long vessel as radiator. Beam particle momentum, direction and time are measured with \SI{0.15}{\GeVc}, \SI{16}{\micro\radian} and \SI{100}{ps} resolutions by a silicon pixel spectrometer consisting of four stations (GTK0,1,2,3) and four dipole magnets. A toroidal muon sweeper, called scraper (SCR), is installed downstream of GTK2. A \SI{1.2}{\m} thick steel collimator (COL) with a \SI{76\times40}{\mm^2} central aperture and \SI{1.7\times1.8}{\m^2} outer dimensions is placed upstream of GTK3 to absorb hadrons from upstream \KP\ decays. Inelastic interactions of beam particles in GTK3 are detected by an array of scintillator bars (CHANTI). A dipole magnet (TRIM5) providing a \SI{90}{\MeVc} horizontal momentum kick is located in front of GTK3. The beam is delivered into a vacuum tank evacuated to a pressure of \SI{10^{-6}}{\milli\bar}, which contains a \SI{75}{\m} long fiducial volume (FV) starting \SI{2.6}{\m} downstream of GTK3. The beam angular spread at the FV entrance is \SI{0.11}{mrad} (rms) in both horizontal and vertical planes. Downstream of the FV, undecayed beam particles continue their path in vacuum.

Momenta of charged particles produced in \KP\ decays are measured by a magnetic spectrometer (STRAW) located in the vacuum tank downstream of the FV. The spectrometer consists of four tracking chambers made of straw tubes, and a large aperture dipole magnet (M), located between the second and third chambers, that provides a horizontal momentum kick of \SI{270}{\MeVc} directed along $-X~(+X)$ for positively (negatively) charged particles. The momentum resolution is $\sigma_p/p = (0.30\oplus 0.005 \cdot p)\%$, with the momentum $p$ expressed in \SI{}{\GeVc}.

A ring-imaging Cherenkov detector (RICH) consisting of a \SI{17.5}{\m} long vessel filled with neon at atmospheric pressure (with a Cherenkov threshold of \SI{12.5}{\GeVc} for pions) provides particle identification, charged particle time measurements with a typical resolution of 70 ps and the trigger time. Two scintillator hodoscopes (CHOD), which include a matrix of tiles  and two planes of slabs arranged in four quadrants located downstream of the RICH, provide trigger signals and time measurements. The tile matrix hodoscope has a time resolution of \SI{1}{ns}, while the slab hodoscope measures time with \SI{200}{\ps} precision.

A 27 radiation length thick liquid krypton (LKr) electromagnetic calorimeter is used as an active neutrino target in this analysis. The calorimeter has a thickness of \SI{127}{cm} and an active volume of \SI{7}{\m^3} segmented in the transverse direction into 13248 projective cells of \SI{2\times2}{\cm^2} size, and provides an energy resolution $\sigma_E/E=(4.8/\sqrt{E}\oplus11/E\oplus0.9)\%$, with $E$ expressed in GeV. To achieve hermetic acceptance for photons emitted in $K^+$ decays in the FV at angles up to \SI{50}{mrad} from the beam axis, the LKr calorimeter is supplemented by annular lead glass detectors (LAV) installed in 12~positions inside and downstream of the vacuum tank, and two lead/scintillator sampling calorimeters (IRC, SAC) located close to the beam axis. An iron/scintillator-strip sampling hadronic calorimeter formed of two modules (MUV1,2) and a muon detector (MUV3), consisting of 148~scintillator tiles located behind an \SI{80}{\cm} thick iron wall, are used for particle identification. The eight smaller tiles of MUV3 adjacent to the beam pipe are referred to as the \textit{inner tiles}, while the remaining 140 regular tiles are called the \textit{outer tiles}.
The four quadrants of MUV3 include only the outer tiles.

 The data sample is obtained from $3.2\times 10^5$ SPS spills collected in 2022, with a typical beam intensity of $3.0\times 10^{12}$ protons per spill. This value corresponds on average to a \SI{600}{MHz} instantaneous beam particle rate at the FV entrance, and a \SI{4.5}{MHz} \KP\ decay rate in the FV. 

\begin{table}[]
    \centering
    \caption{L0 trigger conditions relevant to this study.}  
    \vspace{2pt}
    \begin{tabular}{rl}\hline
    \textbf{L0 Condition}   & \textbf{Requirements} \\\hline
    \texttt{RICH}           & at least two signals in RICH\\
    \texttt{RICH16}         & signals in at least 16 groups of eight adjacent PMTs in the RICH\\
    \texttt{Q1}             & at least one signal in the tile CHOD \\ 
    \texttt{Q2}             & at least one signal in each of two different quadrants of  the tile CHOD\\
    \texttt{UTMC}           & fewer than five signals in the tile CHOD\\
    \texttt{E5}             & at least \SI{5}{GeV} deposited in the LKr\\
    \texttt{MOQX}           & at least one signal in each of two diagonally-opposite quadrants of MUV3\\
    \hline
    \end{tabular}

    \label{tab:L0Conditions}
\end{table}
 
 A two-level trigger system is used, comprising a hardware low level trigger (L0) and a software
high level trigger (L1)~\cite{hep-ph_NA62_2023}. The main trigger line is dedicated to the \KPiNuNuP\ study. Other trigger lines, based on combinations of individual detector conditions, operate concurrently. The L0 trigger conditions relevant to this study are described in \autoref{tab:L0Conditions}.

The signal sample is collected using the Neutrino trigger line. The L0 conditions are
\mbox{\texttt{RICH} $\cdot$ $\overline{\mbox{\texttt{RICH16}}}$ $\cdot$ \texttt{UTMC} $\cdot$ $\overline{\mbox{\texttt{Q2}}}$ $\cdot$ \texttt{E5} $\cdot$ \texttt{MOQX}}. The L1 conditions require \KP\ identification in KTAG, a positively charged track originating in the FV with momentum below \SI{65}{\GeVc} and no activity in LAV. This trigger line selects \KMuNuP\ decays followed by a charged-current neutrino interaction in the LKr calorimeter which produces a hadronic shower and a \MuM.

A normalisation sample of \KMuNuP\ events is used to determine the number of kaon decays and estimate the corresponding number of expected signal events.
This sample is collected using the Minimum bias trigger line (\texttt{RICH} $\cdot$ \texttt{Q1}) downscaled by a factor of 600; no conditions are applied at L1.

Monte Carlo (MC) simulations of particle interactions with the detector and its response are performed using a software package based on the \geant~\cite{hep-ph_AllisonEtAl_2016} and \genie~\cite{hep-ph_AndreopoulosEtAl_2010,hep-ph_GENIE_2021,hep-ph_GENIE_2022,hep-ph_GENIE_2022a} toolkits (with tune \texttt{G18\_10a\_02\_11b}).

\section{Event selection}
\label{sec:evtsel}
The event selection consists of two sets of criteria. The \textit{common selection} is applied to both signal and normalisation samples, and is designed to select \KMuNuP\ decays with a single positively charged track originating in the FV and identified as a muon. The \textit{interaction selection} is designed to select the signal sample comprising \KMuNuP\ decays followed by a neutrino interaction in the \LKr\ calorimeter.
The signal region defined in the following is kept masked until the completion of the background evaluation.

\subsection{Common selection}

\paragraph{Trigger condition} Events are required to satisfy all RICH and CHOD L0 conditions 
used in the Neutrino and Minimum bias trigger lines (\texttt{RICH} $\cdot$ $\overline{\mbox{\texttt{RICH16}}}$ $\cdot$ \texttt{Q1} $\cdot$ $\overline{\mbox{\texttt{Q2}}}$ $\cdot$ \texttt{UTMC}). 

\paragraph{Positively charged muon} 
A positively charged track must be reconstructed in the STRAW spectrometer with momentum below \SI{65}{\GeVc}. Its extrapolated positions at the CHOD, RICH, LKr, MUV1, MUV2 and MUV3 front planes should be within the respective geometrical acceptances. 
Spatial association of in-time signals in these detectors is required. The RICH time defines the track time. Muon identification is performed as follows:
\begin{itemize}\itemsep0em 
    \item both \textit{track-seeded ring} and \textit{single ring} algorithms exploiting RICH signals and track properties~\cite{NA62:2020fhy} must favour the muon hypothesis;   \item a MUV3 signal must be associated with the extrapolated track position.
\end{itemize}

\paragraph{Parent kaon}
A \KP\ candidate is defined as a beam track reconstructed in GTK in time with a KTAG signal. The \KP\ candidate with the KTAG time closest to the \MuP\ track time defines the parent kaon. Inelastic kaon interactions in GTK3 are excluded by applying the following criteria: no CHANTI signal in time with the \MuP\ track; no other \KP\ candidate forming a vertex with the \MuP\ track at a $Z$ coordinate consistent with the GTK3 position (\SI{100-105} {\meter}).

\paragraph{Decay vertex} 
The mid-point of the smallest segment between the \MuP\ and \KP\ tracks defines the decay vertex. The closest distance of approach of these tracks must be smaller than \SI{4}{mm}. The $Z$ coordinate of the vertex must be in the FV range \SI{105-170}{\meter}. 

\begin{figure}[ht]
    \centering
    \includegraphics[width=0.47\linewidth]{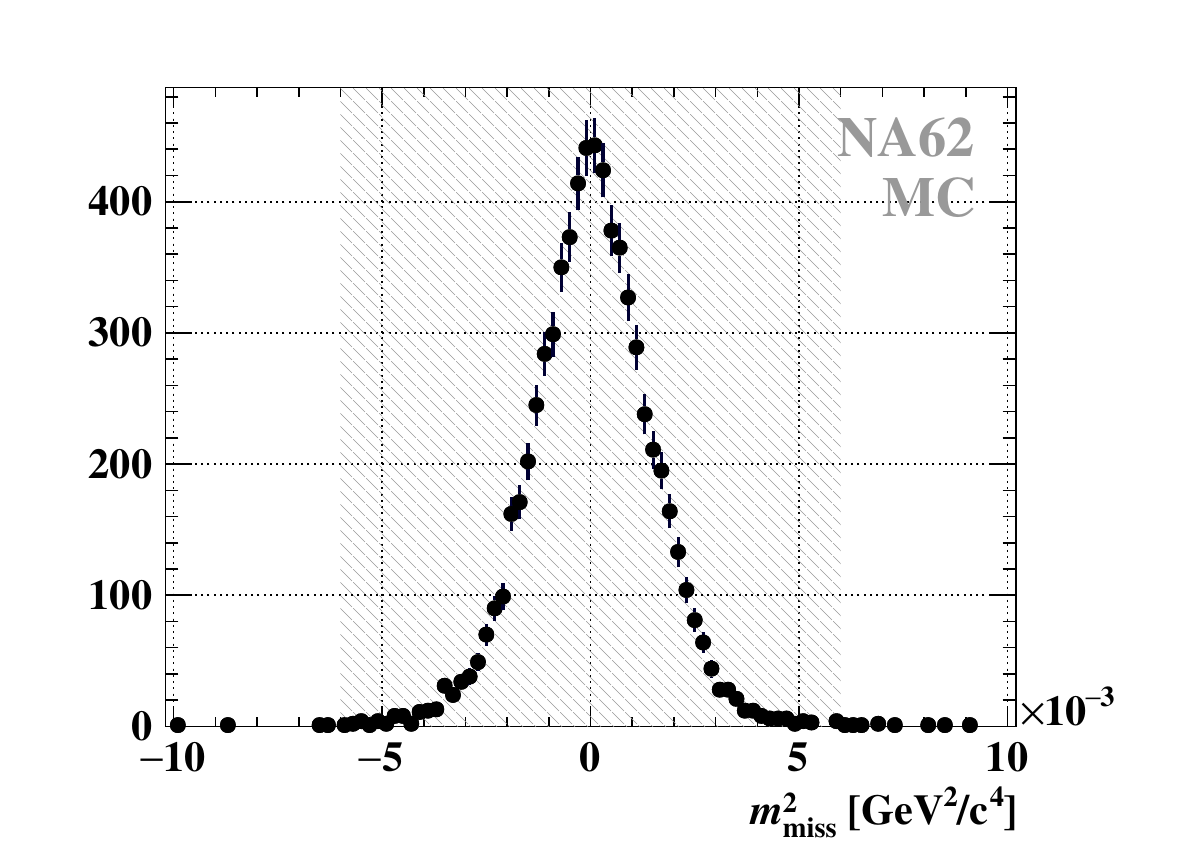}
    \includegraphics[width=0.47\linewidth]{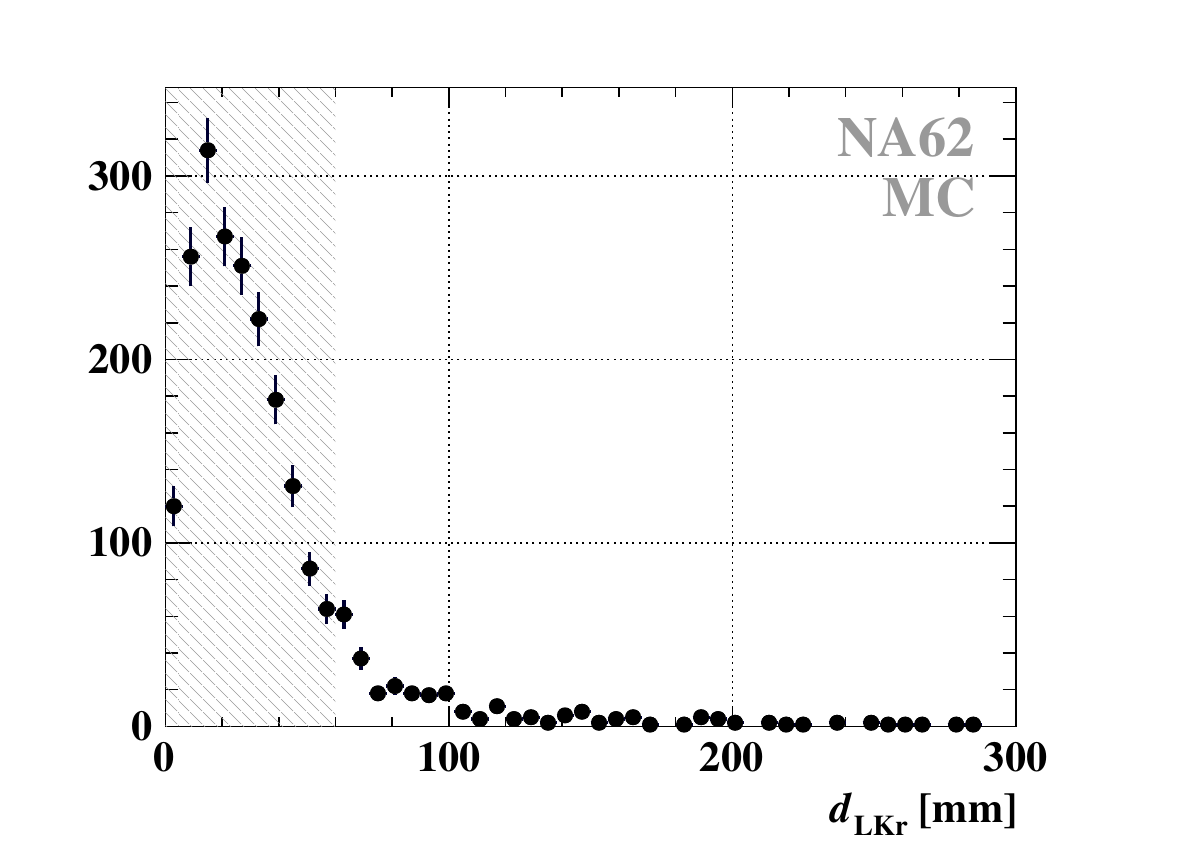}
    \caption{Left: distribution of the reconstructed squared missing mass $\SqMmiss$ for simulated \KMuNuP\ decays. Right: distribution of the distance $d_{\textrm{\LKr}}$ between the extrapolated position of the neutrino candidate at \LKr\ front plane and the associated interaction position. The shaded areas represent the signal region.}
    \label{fig:candidate}
\end{figure}

\paragraph{Neutrino candidate} 
The neutrino candidate is defined as the neutral particle originating from the decay vertex with 4-momentum equal to $P_{\KP} - P_{\MuP}$, where $P_{\KP}$ and $P_{\MuP}$ are the \KP\ and \MuP\ 4-momenta under the assumption of \KP\ and \MuP\ masses, respectively. The squared missing mass, $\SqMmiss = \left( P_{\KP} - P_{\MuP}\right)^2$ must be consistent with zero, $|\SqMmiss|< \SI{0.006}{\mmGeV}$, as shown in \autoref{fig:candidate}-left.
The  extrapolated position of the neutrino candidate at the \LKr\ front plane 
must be within the \LKr\ geometrical acceptance and at least \SI{250}{mm} from the beam axis.
The  extrapolated position at the MUV3 front plane must be at least \SI{300}{mm} from the beam axis. The distance between the extrapolated positions of the neutrino and the \MuP\ candidates should be larger than \SI{200}{mm} at the \LKr\ front plane.

\paragraph{Extra activity}

In-time activity related to other beam tracks or \KP\  decays is excluded by requiring:
\begin{itemize}\itemsep0em 
\item no more than three beam kaons based on KTAG signals detected within \SI{8}{ns} of the RICH time;
\item fewer than five beam tracks in GTK within \SI{0.6}{ns} of the RICH time;
\item no partially reconstructed STRAW tracks pointing to the decay vertex;
\item no activity in LAV downstream of the vertex $Z$ position within \SI{3}{ns} of the RICH time;
\item no more than one signal associated to the \MuP\ in each CHOD hodoscope;
\item no signals in LKr, MUV1, and MUV2 within  \SI{250}{mm}, \SI{200}{mm} and  \SI{300}{mm} of the beam axis, respectively;
\item no signals in SAC, IRC.
\end{itemize}

\subsection{Interaction selection}
\paragraph{Neutrino interaction}
Neutrino interaction signals should be reconstructed in LKr, MUV1, MUV2 and MUV3, matching in time signals from the \MuP\ track from the \KP\ decay.
The distance, $d_{\textrm{\LKr}}$, between the extrapolated position of the neutrino candidate at the \LKr\ front plane and the associated \LKr\ signal position, should be less than \SI{60}{mm}, as shown in \autoref{fig:candidate}-right. 
For kinematic reasons, neutrino interactions are not expected in the LKr inner region centered on the $Z$ axis with a diameter \SI{500}{mm}, extended on the $X > 0$ side to a half-ellipse of \SI{500}{mm} (\SI{900}{mm}) minor (major) axes along $Y(X)$.
Except for the signals associated with the neutrino candidate and the \MuP, no in-time \LKr\ signal should be reconstructed. 
The MUV1 and MUV2 signals should be spatially consistent with the neutrino candidate interaction in LKr.
The  MUV3 signals associated with the neutrino candidate and the \MuP\ should be separated by at least \SI{440}{mm} and located in diagonally opposite quadrants.

\paragraph{Interaction energy}
The LKr energy deposit associated with the neutrino candidate should be larger than \SI{5}{GeV}, to match the \texttt{E5} condition at L0, and between 5\% and 50\% of $E_\nu$, the neutrino energy reconstructed kinematically from the \KMuNuP\ decay. The total in-time energy in MUV1 and MUV2 should be larger than \SI{8}{GeV}. The sum of the energy deposits associated with the interaction in LKr, MUV1 and MUV2 should be smaller than $E_\nu$.
As the hadronic shower is expected to be contained in  LKr and MUV1, the energy deposit associated with the interaction in MUV2 is required to be lower than 10\% of the sum of the associated energy deposits in \LKr, MUV1 and MUV2.

The production angle of the \MuM\ increases as a function of the neutrino energy fraction transferred to the hadronic shower. The distance, $d_{\textrm{MUV3}}$, between the extrapolated position of the neutrino candidate at MUV3 front plane and the MUV3 signal position is expected to decrease linearly with the inelasticity $y = (E_\nu - E_\textrm{LKr} - E_\textrm{MUV1})/E_\nu$, where $E_\textrm{LKr}$ and $E_\textrm{MUV1}$ are the energy deposits associated with the neutrino candidate in \LKr\ and MUV1. As a result, $d_{\textrm{MUV3}}$ is required to satisfy the condition $\SI{880}{mm}  < d_{\textrm{MUV3}} + \SI{880}{mm} \times y < \SI{1300}{mm}$. 

\paragraph{Interaction topology} 
The Fermi motion of the target nucleon being negligible at the mean neutrino energy of \SI{40}{\GeV}, the neutrino and its interaction products, the hadronic shower and the \MuM, should be coplanar. Using a cylindrical coordinate system with the neutrino candidate direction defining the longitudinal axis, the azimuthal angle, $\varphi$, between the LKr signal associated with the neutrino candidate and the MUV3 signal associated with the \MuM\ must be compatible with $\pi$ radians, and is required to satisfy $|\varphi-\pi|<\SI{1.44}{\radian}$.

\section{Background estimate}
\label{sec:bkg}
The background events originate from (A) \KMuNuP\ decays with unrelated in-time activity mimicking a neutrino interaction, and (B) mis-reconstructed non-\KMuNuP\ decays.

A data-driven method is used to estimate the contribution of background events by modelling their observed distribution in the regions $|\SqMmiss| \geq \SI{0.006}{\mmGeV}$ or $d_\LKr \geq \SI{60}{mm}$ and integrating the model in the signal region.

\begin{figure}[t]
	\begin{center}
		\newlength{\mylength}
		\setlength{\mylength}{\heightof{\includegraphics[width=0.49\textwidth]{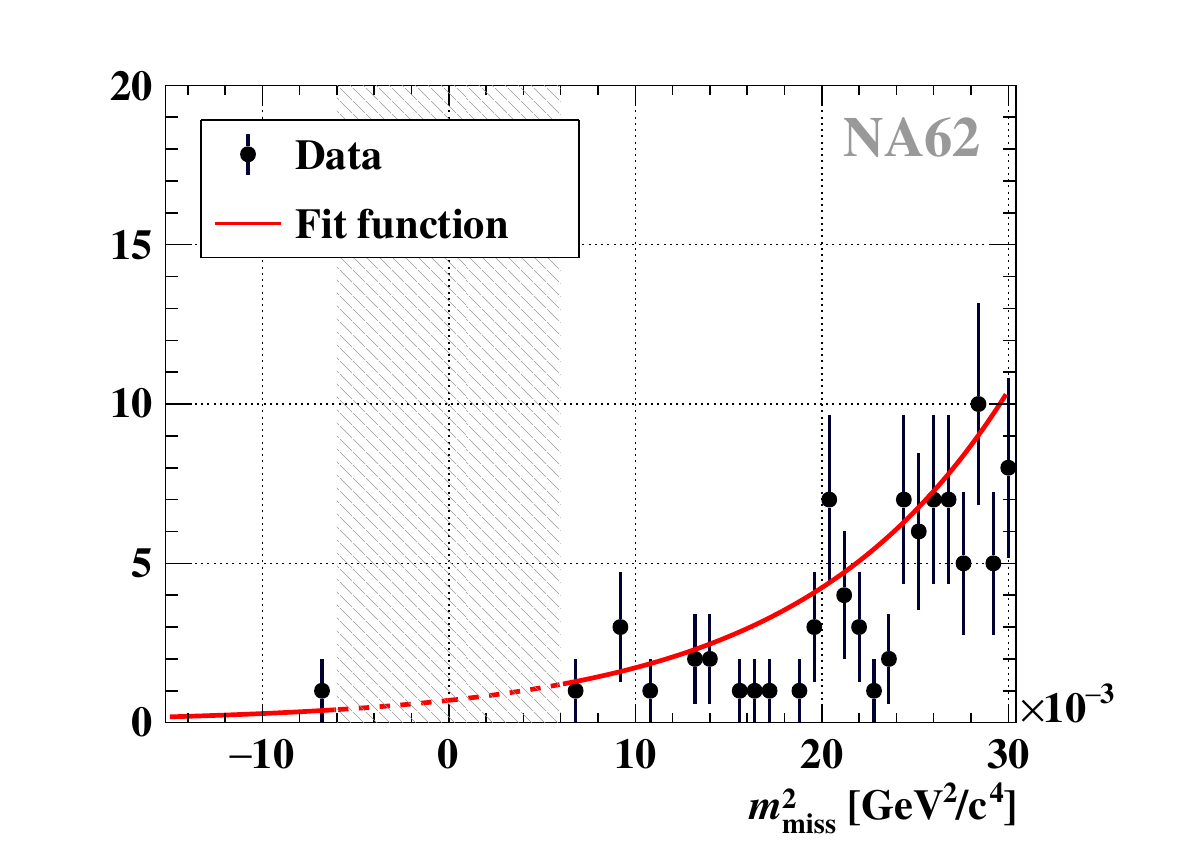}}}
		\resizebox{!}{\mylength}{\includegraphics{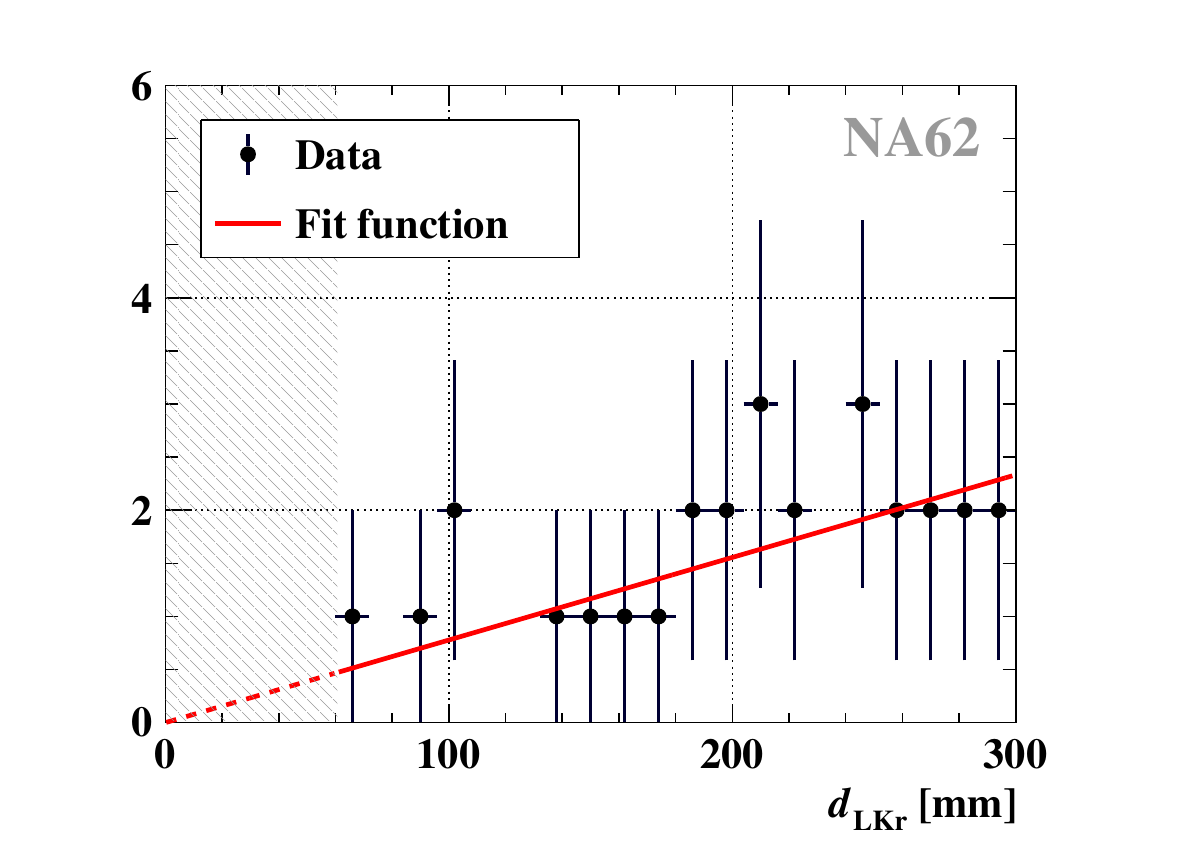}}
	\resizebox{!}{\mylength}{\includegraphics{MisRecoBkgFit.pdf}}
\end{center}
	\vspace{-9mm}
	\caption{Left: $d_\LKr$ distribution of the selected events in the background A sample, fitted with a linear function.
Right: $m^2_{\rm miss}$ distribution of the selected events in the background B sample, fitted with an exponential function. The shaded areas represent the masked signal region.
}
	\label{fig:Bkg}
\end{figure}

{\bf Background A} is studied using a sample satisfying the signal selection but inverting the $d_\LKr$ condition and removing the requirements on $\varphi$ and on $d_{\textrm{MUV3}}$ to increase the sample size.
Assuming that the accidental activity is uniform over the LKr front plane, the number of background events per unit of $d_\LKr$ is proportional to $d_\LKr$.
The $d_\LKr$ distribution of the selected sample is fitted with a linear function
from \SIrange{60}{300}{mm} as shown in \autoref{fig:Bkg}-left. The expected number of events in the signal region is obtained as the integral of the fit function in this region, scaled by the efficiency of the removed selection criteria, which is measured in the side-bands.
The contribution is found to be \num{0.030~ ^{+0.041}_{-0.023} \big|_\stat \pm 0.004_\syst}. The statistical uncertainties account for those of the fit parameter and the scale factor. The systematic uncertainty accounts for the choice of the fit range, and is computed as the difference between the scaled integrals using the fit ranges \SI{[60, 300]}{mm} and \SI{[60, 180]}{mm}.

{\bf Background B} is similarly studied by inverting the \SqMmiss\ requirement and removing the interaction energy and topology conditions to increase the sample size. The $m^2_{\rm miss}$ distribution of the selected sample is fitted from \num{-0.015} to \SI{0.03}{\mmGeV} with an exponential function, excluding the signal region $|\SqMmiss| < $\SI{0.006}{\mmGeV}, as shown in \autoref{fig:Bkg}-right. 
The integration of the fit function in the signal region, after scaling by the measured efficiency of the removed criteria, gives an estimated contribution of 
\num{0.0039~^{+0.0060}_{-0.0034}\big|_\stat \pm 0.0007_\syst}. The statistical uncertainties account for those of the fit parameters and scale factor. The systematic uncertainty accounts for the choice of the fit function and is computed as the difference between the scaled integrals using  exponential and quadratic functions.

The total number of expected background events in the signal region, dominated by background A, is $0.034~^{+0.041}_{-0.023}\big|_\stat \pm 0.004_\syst$.

\section{Expected signal}  
\label{sec:signal}

The expected number of signal events, $N_\sig$, is obtained from the number of observed normalisation events, $N_\norm =2.22 \times 10^8$: 
\begin{equation*}
	N_\sig = N_\norm \cdot D \cdot 
    P^{\rm{int}} \cdot
	\frac{\varepsilon_\sig}{\varepsilon_\norm},
	\label{eq:SigExp1}
\end{equation*}
where $D = 600$ is the downscaling factor of the L0 Minimum bias trigger line used to collect the normalisation events, 
$P^{\rm{int}}$
is the average neutrino interaction probability for the selected signal sample, and 
$\varepsilon_\sig$ ($\varepsilon_\norm$) is the selection and trigger efficiency of the signal (normalisation) sample. 

Given the \NuMu\ charged-current interaction cross section (\SI{0.68\times10^{-38}}{cm^{2}\GeV^{-1}}~\cite{hep-ph_FormaggioEtAl_2012}), the mean neutrino energy (\SI{40}{\GeV}) and the thickness of the \LKr\ calorimeter (\SI{127}{cm}), the interaction probability, $P^{\rm{int}}$, is expected to be a few $10^{-11}$. A more precise value is obtained using simulations by GENIE including elaborated models of cross sections. The neutrino from each simulated \KMuNuP\ decay is transferred to GENIE which calculates the neutrino interaction probability in the \LKr\ calorimeter, generates the interaction position and the final state particles, and returns this information to the NA62 simulation software. The average interaction probability for the selected signal events is found to be $P^{\rm{int}} = (5.6 \pm 0.1_\syst)\times 10^{-11}$, where the uncertainty is due to the precision of the neutrino cross-section measurements.

Removing the trigger components common to the signal and normalisation selections, the efficiency ratio becomes
\begin{equation*}
\label{eq:SigExp2}
\frac{\varepsilon_\sig}{\varepsilon_\norm} = 
\varepsilon^{\texttt{int}}\cdot \varepsilon^{\texttt{MOQX}} \cdot
\varepsilon^{\texttt{E5}} \cdot
\varepsilon^{\texttt{L1}},
\end{equation*}
where $\varepsilon^{\texttt{int}},\varepsilon^{\texttt{MOQX}}, \varepsilon^{\texttt{E5}}$, and $\varepsilon^{\texttt{L1}}$ are the efficiencies of the interaction selection and trigger conditions applied only in the signal selection. 

The interaction selection efficiency is $\varepsilon^{\texttt{int}} = 0.0367\pm 0.0022_{\stat} \pm 0.0013_{\syst}$, as
evaluated using simulations and correcting for the random loss induced by non-simulated extra activity in the detectors. This loss is estimated from data as the fraction of events rejected when requiring in-time activity in LKr and MUV3 in the normalisation sample.

The \texttt{MOQX} trigger efficiency is measured using a sample of $\KP\to\PiP\PiP\PiM$ decays where two pions of opposite charge decay in flight as $\PiMuNu$. The sample is collected with a trigger line based only on RICH and CHOD signals and therefore independent of MUV3. Events are required to have two reconstructed tracks associated with MUV3 signals located in diagonally opposite outer quadrants. The \texttt{MOQX} efficiency is computed as a function of the MUV3 tile pair. The efficiency $\varepsilon^{\texttt{MOQX}} = 0.976 \pm 0.007_\stat$ is obtained by weighting the tile-pair efficiencies by the tile-pair distribution obtained from signal simulation.

The \texttt{E5} trigger efficiency is evaluated using a sample of \KPiPi\ decays where the \PiP\ produces a hadronic shower in LKr and the photons from the \PiZ\ decay are detected in the LAV. 
The sample comprises events collected by trigger lines that do not rely on \LKr\ signals. Events are required to have no activity in the \LKr\ inner region, IRC and SAC and exactly one \LKr\ signal in time with the \PiP\ track, ensuring a purely hadronic energy deposit in \LKr. The \texttt{E5} efficiency is measured as a function of the in-time energy deposit in the LKr.
The efficiency $\varepsilon^{\texttt{E5}} = 0.847 \pm 0.008_{\stat} \pm 0.003_{\syst}$
is obtained by weighting the measurements in individual energy bins by the energy distribution of selected signal events obtained from simulations. The systematic uncertainty is obtained by varying conservatively by 10\% the neutrino energy used in the GENIE simulation of the interaction~\cite{hep-ph_MINOS_2010}.

The \texttt{L1} efficiency is found to be  $\varepsilon^{\texttt{L1}} = 0.918 \pm 0.002_\stat$, using a downscaled dataset recorded without \texttt{L1} conditions applied~\cite{hep-ph_NA62_2023}.

The resulting number of expected signal events is $$N_\sig = 0.208 \pm 0.013_\stat \pm 0.009_{\syst},$$
leading to a signal over background ratio of \num{6/1}.
\newpage
\section{Results and prospects}
\label{sec:results}
\begin{figure}[h!t]
	\begin{center}
		\resizebox{\textwidth}{!}{\includegraphics{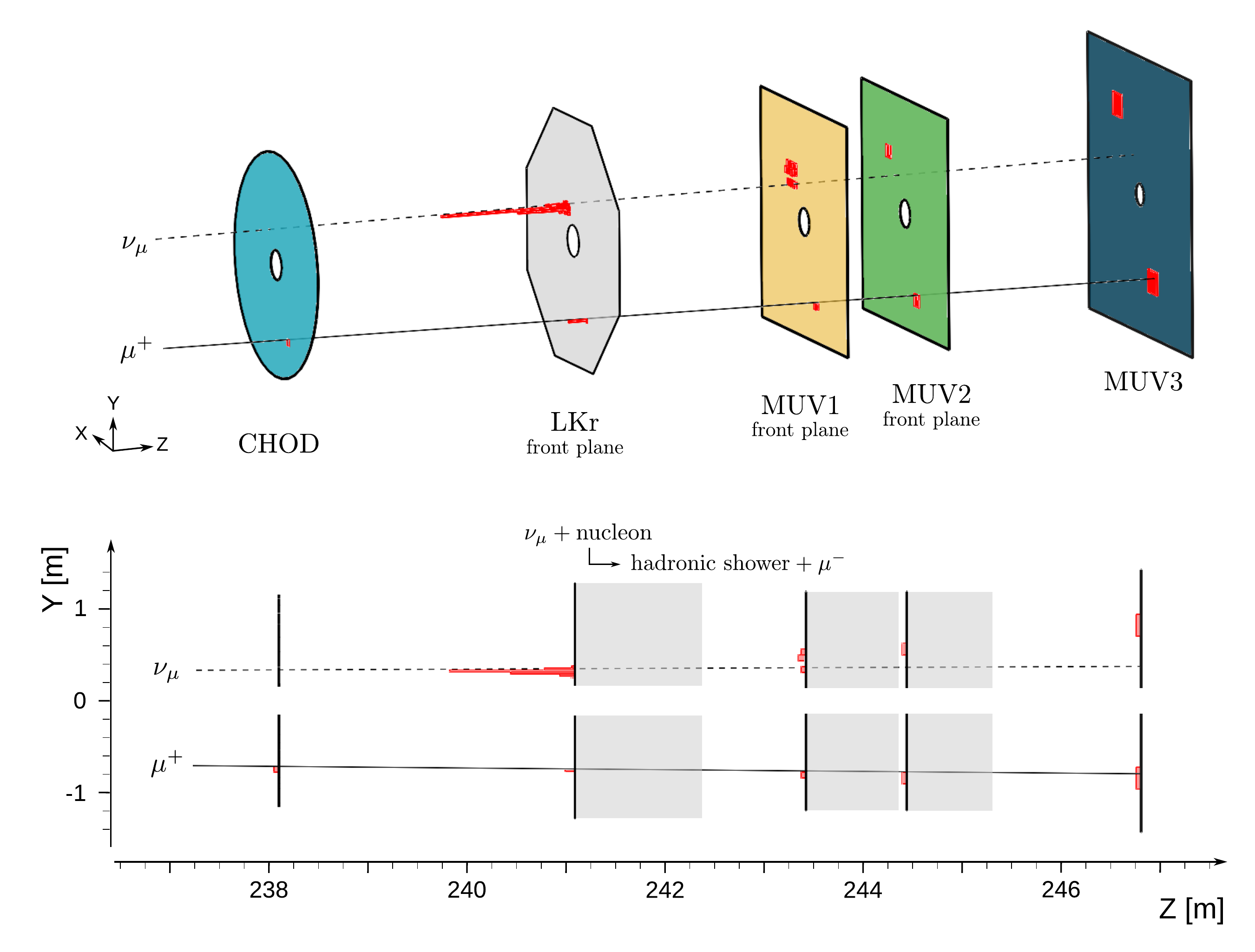}}
	\end{center}
	\vspace{-12mm}
	\caption{Display of the activity recorded in CHOD, LKr, MUV1, MUV2 and MUV3 for the signal candidate. No extra in-time activity is recorded in these detectors. The red histograms represent the activity in each detector projected at the front plane. For LKr, MUV1 and MUV2 the heights correspond to the fraction of the total energy deposited in each cell.  
 The solid line represents the \MuP\ trajectory extrapolated from the STRAW track; the dashed line represents the neutrino trajectory reconstructed using the \KP\ and \MuP\ momenta.
 The shaded areas in the projection represent the detector volumes.
 }
	\label{fig:EvtA}
\end{figure}

After unmasking the signal region, one event is observed,
the main properties of which are reported in \autoref{tab:EvtProp}. A display of the activity in the downstream detectors is shown in \autoref{fig:EvtA}.
Given the total number of expected events $(0.242^{+0.043}_{-0.027}\big|_\stat \pm 0.010_\syst$),
the probability of observing one event in the signal region is \num{19\%} while the probability of observing one background event candidate is \num{3\%}.

The observed event is the first tagged neutrino candidate and demonstrates the feasibility of the neutrino tagging technique. 
The obtained result is an important step toward dedicated tagged neutrino experiments, allowing the collection of large samples of tagged neutrinos, which would open new perspectives for accelerator-based neutrino physics.
The data already collected by NA62 in 2023--2024 can be used to consolidate the result of the study presented here.

\begin{table}[!ht]
\centering
\renewcommand{\arraystretch}{1.1} 
\caption{Properties of the tagged neutrino candidate event. Its energy evaluated from kinematics has a relative uncertainty of 0.34\% which is significantly better than the 10-30\% typically achievable in neutrino detectors.
}
\begin{tabular}{rl}\small
\renewcommand{\arraystretch}{0.8}
\textbf{Property}      & \textbf{Value}\\ 
\hline
\KP\ momentum    & \SI{77.34}{\GeVc} \\ 
\MuP\ momentum   & \SI{25.26}{\GeVc}  \\ 
Decay vertex position $z_{vtx}$  &  \SI{161.2}{m}  \\
Neutrino energy $E_{\nu} = E_{\KP} - E_{\MuP}$        & \SI{52.09}{GeV} \\  
Distance of the LKr signal to the $Z$ axis  & \SI{334.9}{mm} \\
\LKr\ energy associated with the neutrino  $E_\LKr$    & \SI{12.54}{GeV} \\
MUV1 energy associated with the neutrino $E_\mathrm{MUV1}$   & \SI{13.75}{GeV}  \\
MUV2 energy associated with the neutrino  $E_\mathrm{MUV2}$  & \SI{10.74}{GeV}  \\
Inelasticity $y$      & 0.69 \\
Azimuthal angle between the \LKr\ and MUV3 signals $\varphi$   & \SI{3.28}{rad} \\
$d_\LKr$             & \SI{31.4}{mm}  \\\
$d_{\textrm{MUV3}}$  &   \SI{567.9}{mm}\\
$m^2_{\rm miss}$     & \SI{-0.00086}{\mmGeV} \\
\hline
	\end{tabular}	
	\label{tab:EvtProp}
\end{table}

\clearpage

%% file: acknowrun2022_optC_nutag-v3.tex
The authors are grateful to Marco Roda and Costas Andreopoulos from the GENIE Collaboration for their assistance in implementing the neutrino interaction simulation in the NA62 software.
It is a pleasure to express our appreciation to the staff of the CERN laboratory and the technical
staff of the participating laboratories and universities for their efforts in the operation of the
experiment and data processing.

The cost of the experiment and its auxiliary systems was supported by the funding agencies of 
the Collaboration Institutes. We are particularly indebted to: 
F.R.S.-FNRS (Fonds de la Recherche Scientifique - FNRS), under Grants No. 4.4512.10, 1.B.258.20, Belgium;
CECI (Consortium des Equipements de Calcul Intensif), funded by the Fonds de la Recherche Scientifique de Belgique (F.R.S.-FNRS) under Grant No. 2.5020.11 and by the Walloon Region, Belgium;
NSERC (Natural Sciences and Engineering Research Council), funding SAPPJ-2018-0017,  Canada;
MEYS (Ministry of Education, Youth and Sports) funding LM 2018104, Czech Republic;
BMBF (Bundesministerium f\"{u}r Bildung und Forschung), Germany;
INFN  (Istituto Nazionale di Fisica Nucleare),  Italy;
MIUR (Ministero dell'Istruzione, dell'Universit\`a e della Ricerca),  Italy;
CONACyT  (Consejo Nacional de Ciencia y Tecnolog\'{i}a),  Mexico;
IFA (Institute of Atomic Physics) Romanian 
CERN-RO Nr. 06/03.01.2022
and Nucleus Programme PN 19 06 01 04,  Romania;
MESRS  (Ministry of Education, Science, Research and Sport), Slovakia; 
CERN (European Organization for Nuclear Research), Switzerland; 
STFC (Science and Technology Facilities Council), United Kingdom;
NSF (National Science Foundation) Award Numbers 1506088 and 1806430,  U.S.A.;
ERC (European Research Council)  ``UniversaLepto'' advanced grant 268062, ``KaonLepton'' starting grant 336581, Europe.

Individuals have received support from:
Charles University (grants UNCE 24/SCI/016, PRIMUS 23/SCI/025), Ministry of Education, Youth and Sports (project FORTE CZ.02.01.01 /0022-008/0004632), Czech Republic; Czech Science Foundation (grant 23-06770S);   \\
Agence Nationale de la Recherche (grant ANR-19-CE31-0009), France;
Ministero dell'Istruzione, dell'Universit\`a e della Ricerca (MIUR  ``Futuro in ricerca 2012''  grant RBFR12JF2Z, Project GAP), Italy;
the Royal Society  (grants UF100308, UF0758946), United Kingdom;
STFC (Rutherford fellowships ST/J00412X/1, ST/M005798/1), United Kingdom;
ERC (grants 268062,  336581 and  starting grant 802836 ``AxScale'');
EU Horizon 2020 (Marie Sk\l{}odowska-Curie grants 701386, 754496, 842407, 893101, 101023808).

%% file: main.bbl
\begin{thebibliography}{10}
\expandafter\ifx\csname url\endcsname\relax
  \def\url#1{\texttt{#1}}\fi
\expandafter\ifx\csname urlprefix\endcsname\relax\def\urlprefix{URL }\fi
\expandafter\ifx\csname href\endcsname\relax
  \def\href#1#2{#2} \def\path#1{#1}\fi

\bibitem{hep-ph_DUNE_2020}
B.~Abi, et~al., {Long-baseline neutrino oscillation physics potential of the DUNE experiment}, Eur. Phys. J. C 80~(10) (2020) 978.
\newblock \href {http://arxiv.org/abs/2006.16043} {\path{arXiv:2006.16043}}, \href {https://doi.org/10.1140/epjc/s10052-020-08456-z} {\path{doi:10.1140/epjc/s10052-020-08456-z}}.

\bibitem{hep-ph_HyperKamiokande_2018}
K.~Abe, et~al., {Hyper-Kamiokande Design Report}, \url{https://doi.org/10.48550/arXiv.1805.04163} (5 2018).
\newblock \href {http://arxiv.org/abs/1805.04163} {\path{arXiv:1805.04163}}, \href {https://doi.org/10.48550/arxiv.1805.04163} {\path{doi:10.48550/arxiv.1805.04163}}.

\bibitem{hep-ph_ESPP_2020}
{2020 Update of the European Strategy for Particle Physics (Brochure)}, \url{http://cds.cern.ch/record/2721370} (2020).
\newblock \href {https://doi.org/10.17181/CERN.JSC6.W89E} {\path{doi:10.17181/CERN.JSC6.W89E}}.

\bibitem{hep-ph_HuberEtAl_2022}
P.~Huber, et~al., {Snowmass Neutrino Frontier Report}, \url{https://doi.org/10.48550/arXiv.2211.08641} (Nov. 2022).
\newblock \href {https://doi.org/10.48550/ARXIV.2211.08641} {\path{doi:10.48550/ARXIV.2211.08641}}.

\bibitem{hep-ph_HuberEtAl_2008}
P.~Huber, et~al., {On the impact of systematical uncertainties for the CP violation measurement in superbeam experiments}, JHEP 03 (2008) 021.
\newblock \href {https://doi.org/10.1088/1126-6708/2008/03/021} {\path{doi:10.1088/1126-6708/2008/03/021}}.

\bibitem{hep-ph_BrancaEtAl_2021}
A.~Branca, G.~Brunetti, A.~Longhin, M.~Martini, F.~Pupilli, F.~Terranova, {A New Generation of Neutrino Cross Section Experiments: Challenges and Opportunities}, Symmetry 13~(9) (2021) 1625.
\newblock \href {https://doi.org/10.3390/sym13091625} {\path{doi:10.3390/sym13091625}}.

\bibitem{hep-ph_DUNE_2021}
A.~Abed~Abud, et~al., {Deep Underground Neutrino Experiment (DUNE) Near Detector Conceptual Design Report}, Instruments 5~(4) (2021) 31.
\newblock \href {https://doi.org/10.3390/instruments5040031} {\path{doi:10.3390/instruments5040031}}.

\bibitem{hep-ph_Pontecorvo_1979}
B.~Pontecorvo, {Tagging direct neutrinos: a first step to neutrino tagging}, Lett. Nuovo Cim. 25 (1979) 257--259.
\newblock \href {https://doi.org/10.1007/BF02813638} {\path{doi:10.1007/BF02813638}}.

\bibitem{hep-ph_Nedyalkov_1984}
I.~P. Nedyalkov, \href{http://inis.jinr.ru/sl/NTBLIB/JINR-E1-84-515.pdf}{Single spectrometer station for neutrino tagging}, \url{http://inis.jinr.ru/sl/NTBLIB/JINR-E1-84-515.pdf} (7 1984).
\newline\urlprefix\url{http://inis.jinr.ru/sl/NTBLIB/JINR-E1-84-515.pdf}

\bibitem{hep-ph_Bohm_1987}
G.~Bohm, \href{https://inis.iaea.org/search/21064823}{{Project of a tagged neutrino facility at Serpukhov}}, in: Physics at Future Accelerators. Proceedings, 10th Warsaw Symposium on Elementary Particle Physics, Kazimierz, Poland, 1987, 1987.
\newline\urlprefix\url{https://inis.iaea.org/search/21064823}

\bibitem{hep-ph_Bernstein_1993}
R.~Bernstein, {Neutrino physics at Fermilab in the 1990’s}, Nuclear Physics B - Proceedings Supplements 31 (1993) 255--261.
\newblock \href {https://doi.org/10.1016/0920-5632(93)90143-t} {\path{doi:10.1016/0920-5632(93)90143-t}}.

\bibitem{hep-ph_PERRIN-TERRIN_2022}
M.~Perrin-Terrin, Neutrino tagging: a new tool for accelerator based neutrino experiments, Eur. Phys. J. C 82~(5) (2022) 465.
\newblock \href {https://doi.org/10.1140/epjc/s10052-022-10397-8} {\path{doi:10.1140/epjc/s10052-022-10397-8}}.

\bibitem{hep-ph_FriedlandEtAl_2019}
A.~Friedland, S.~W. Li, Understanding the energy resolution of liquid argon neutrino detectors, Phys. Rev. D 99~(3) (2019) 036009.
\newblock \href {http://arxiv.org/abs/1811.06159} {\path{arXiv:1811.06159}}, \href {https://doi.org/10.1103/PhysRevD.99.036009} {\path{doi:10.1103/PhysRevD.99.036009}}.

\bibitem{hep-ph_ENUBET_2023}
F.~Acerbi, et~al., {Design and performance of the ENUBET monitored neutrino beam}, The European Physical Journal C 83~(10) (Oct. 2023).
\newblock \href {https://doi.org/10.1140/epjc/s10052-023-12116-3} {\path{doi:10.1140/epjc/s10052-023-12116-3}}.

\bibitem{hep-ph_Baratto-RoldanEtAl_2024}
A.~Baratto-Roldán, M.~Perrin-Terrin, E.~Parozzi, M.~Jebramcik, N.~Charitonidis, Nutag: a proof-of-concept study for a long-baseline neutrino beam, Eur. Phys. J. C 84 (10 2024).
\newblock \href {https://doi.org/10.1140/epjc/s10052-024-13324-1} {\path{doi:10.1140/epjc/s10052-024-13324-1}}.

\bibitem{hep-ph_BorgatoEtAl_2023}
F.~Borgato, et~al., \href{https://www.frontiersin.org/articles/10.3389/ fphy.2023.1117575}{Charged-particle timing with 10 ps accuracy using {TimeSPOT} {3D} trench-type silicon pixels}, Frontiers in Physics 11 (2023).
\newline\urlprefix\url{https://www.frontiersin.org/articles/10.3389/ fphy.2023.1117575}

\bibitem{CortinaGil2021}
E.~Cortina~Gil, et~al., {Measurement of the very rare $K^+\rightarrow\pi^+\nu\bar{\nu}$ decay}, JHEP 06 (2021) 93.
\newblock \href {http://arxiv.org/abs/2103.15389} {\path{arXiv:2103.15389}}, \href {https://doi.org/10.1007/JHEP06(2021)093} {\path{doi:10.1007/JHEP06(2021)093}}.

\bibitem{hep-ph_AglieriRinellaEtAl_2019}
G.~Aglieri~Rinella, et~al., {The NA62 GigaTracKer: a low mass high intensity beam 4D tracker with 65 ps time resolution on tracks}, JINST 14 (2019) P07010.
\newblock \href {http://arxiv.org/abs/1904.12837} {\path{arXiv:1904.12837}}, \href {https://doi.org/10.1088/1748-0221/14/07/P07010} {\path{doi:10.1088/1748-0221/14/07/P07010}}.

\bibitem{hep-ph_NA62_2017}
E.~Cortina~Gil, et~al., {The Beam and detector of the NA62 experiment at CERN}, JINST 12~(05) (2017) P05025.
\newblock \href {http://arxiv.org/abs/1703.08501} {\path{arXiv:1703.08501}}, \href {https://doi.org/10.1088/1748-0221/12/05/P05025} {\path{doi:10.1088/1748-0221/12/05/P05025}}.

\bibitem{hep-ph_NA62_2023}
E.~Cortina~Gil, et~al., {Performance of the NA62 trigger system}, JHEP 03 (2023) 122.
\newblock \href {https://doi.org/10.1007/JHEP03(2023)122} {\path{doi:10.1007/JHEP03(2023)122}}.

\bibitem{hep-ph_AllisonEtAl_2016}
J.~Allison, et~al., {Recent developments in Geant4}, Nucl. Instrum. Meth. A 835 (2016) 186--225.
\newblock \href {https://doi.org/10.1016/j.nima.2016.06.125} {\path{doi:10.1016/j.nima.2016.06.125}}.

\bibitem{hep-ph_AndreopoulosEtAl_2010}
C.~Andreopoulos, et~al., {The GENIE Neutrino Monte Carlo Generator}, Nucl. Instrum. Meth. A 614 (2010) 87--104.
\newblock \href {https://doi.org/10.1016/j.nima.2009.12.009} {\path{doi:10.1016/j.nima.2009.12.009}}.

\bibitem{hep-ph_GENIE_2021}
J.~Tena-Vidal, et~al., {Neutrino-nucleon cross-section model tuning in GENIE v3}, Phys. Rev. D 104~(7) (2021) 072009.
\newblock \href {https://doi.org/10.1103/PhysRevD.104.072009} {\path{doi:10.1103/PhysRevD.104.072009}}.

\bibitem{hep-ph_GENIE_2022}
J.~Tena-Vidal, et~al., {Neutrino-nucleus CC0$\pi$ cross-section tuning in GENIE v3}, Phys. Rev. D 106~(11) (2022) 112001.
\newblock \href {https://doi.org/10.1103/PhysRevD.106.112001} {\path{doi:10.1103/PhysRevD.106.112001}}.

\bibitem{hep-ph_GENIE_2022a}
J.~Tena-Vidal, et~al., {Hadronization model tuning in GENIE v3}, Phys. Rev. D 105~(1) (2022) 012009.
\newblock \href {https://doi.org/10.1103/PhysRevD.105.012009} {\path{doi:10.1103/PhysRevD.105.012009}}.

\bibitem{NA62:2020fhy}
E.~Cortina~Gil, et~al., {An investigation of the very rare $ {K}^{+}\to {\pi}^{+}\nu \overline{\nu} $ decay}, JHEP 11 (2020) 042.
\newblock \href {http://arxiv.org/abs/2007.08218} {\path{arXiv:2007.08218}}, \href {https://doi.org/10.1007/JHEP11(2020)042} {\path{doi:10.1007/JHEP11(2020)042}}.

\bibitem{hep-ph_FormaggioEtAl_2012}
J.~A. Formaggio, G.~P. Zeller, {From eV to EeV: Neutrino Cross Sections Across Energy Scales}, Rev. Mod. Phys. 84 (2012) 1307--1341.
\newblock \href {https://doi.org/10.1103/RevModPhys.84.1307} {\path{doi:10.1103/RevModPhys.84.1307}}.

\bibitem{hep-ph_MINOS_2010}
P.~Adamson, et~al., Neutrino and antineutrino inclusive charged-current cross section measurements with the minos near detector, Phys. Rev. D 81 (2010) 072002.
\newblock \href {http://arxiv.org/abs/0910.2201} {\path{arXiv:0910.2201}}, \href {https://doi.org/10.1103/PhysRevD.81.072002} {\path{doi:10.1103/PhysRevD.81.072002}}.

\end{thebibliography}
